\documentstyle[epsf,eqsecnum,floats,preprint,aps]{revtex}

\tighten

\input epsf
\begin{document}

\newcommand{\be}{\begin{equation}}
\newcommand{\ee}{\end{equation}}
\newcommand{\bea}{\begin{eqnarray}}
\newcommand{\eea}{\end{eqnarray}}
\newcommand{\PSbox}[3]{\mbox{\rule{0in}{#3}\includegraphics{#1}\hspace{#2}}}

\overfullrule=0pt
\def\d{\partial}
\def\e{\epsilon}
\def\gsim{\mathrel{\raise.3ex\hbox{$>$}\mkern-14mu
              \lower0.6ex\hbox{$\sim$}}}
\def\lsim{\mathrel{\raise.3ex\hbox{$<$}\mkern-14mu
              \lower0.6ex\hbox{$\sim$}}}

\rightline{astro-ph/0303268}
\vskip 4cm

\setcounter{footnote}{0}

\begin{center}
\large{\bf 
A post-WMAP perspective on inflation
}
\ \\
\ \\
\normalsize{Arthur Lue,\footnote{E-mail:  lue@bifur.cwru.edu}
  Glenn D. Starkman,\footnote{E-mail:  glenn.starkman@cwru.edu}
  and Tanmay Vachaspati\footnote{E-mail:  tanmay@monopole.cwru.edu}
}
\ \\
\ \\
\small{\em Center for Education and Research in Cosmology and Astrophysics\\
Department of Physics\\
Case Western Reserve University \\
Cleveland, OH 44106--7079}

\end{center}

\begin{abstract}

\noindent
Recent results from the Wilkinson Microwave Anisotropy Probe have been
called a corroboration, or even a confirmation, of inflation.  Yet, the
results include features that require, at least, a significant
distortion of what is usually meant by inflation.  At the same time,
critics have leveled the charge that inflation is an arbitrarily
pliable theory and is therefore beyond proof or disproof.  This
startling dissonance in attitudes toward inflation seems to have
grown out of the lack of a clear framework with which to evaluate the
inflationary paradigm.  In this rhetorical pamphlet we reexamine the
inflationary paradigm, attempt to articulate explicitly how the
paradigm and its descendant models are falsifiable, and make a sober
assessment of the successes and failures of inflation.
\end{abstract}

\setcounter{page}{0}
\thispagestyle{empty}
\maketitle

\eject

\vfill

\baselineskip 18pt plus 2pt minus 2pt

\section{Introduction}

The dawn of the 21st century has indeed yielded the promised golden
age of modern cosmology.  The wealth of observational data from both
satellites and ground-based surveys provide an increasingly refined
set of tools for probing and criticizing the increasingly coherent
theoretical framework of the standard cosmological model: a hot big
bang evolution of a universe filled with cold dark matter, with an
early period of inflation that provides flatness and homogeneity in
the observable Universe and which, at the same time, provides the
source of primordial density fluctuations from which all observed
structure evolved.

The recent results from the first year of data from NASA's Wilkinson
Microwave Anisotropy Probe (WMAP) are a remarkable accomplishment, a
tour de force of fantastic and careful analysis.  The NASA press
conference announcing the results of WMAP claimed that the data
provides a confirmation, or at least a corroboration, of the
inflationary paradigm.  This last phrase, ``the inflationary
paradigm,'' has given rise to considerable angst amongst cosmologists.
The mantra that inflation is not a theory, rather it is a paradigm,
has been used by enthusiasts and detractors alike.  Proponents claim
that inflation is a simple but powerful environment where one can
study a large variety of models and answer a host of questions.
Critics respond by questioning whether inflation is really science
under those circumstances, and assert that inflation, as a paradigm
rather than a theory, can be engineered to provide whatever result is
necessary.  Indeed, the claim that WMAP corroborates inflation merely
confirmed the worst fears of inflation detractors: how can one confirm
a paradigm that can never be disproven?

The cosmology community must surely demand that the pillars of its
standard theoretical framework have firm foundations in scientific
principles: providing explanations of known information, offering new
predictions, and subjecting itself to falsification.  Is inflation
good science?  Perhaps -- what is clear is that the criticisms of
inflation as a scientific paradigm are not entirely unjustified.  We
wish to lay out a set of sober thoughts regarding inflation, both pro
and con.  Little in this discussion will be new.  Consider this a
rhetorical pamphlet rather than a paper, one where we attempt to
collect and organize ideas that many have expressed, to give voice to
the frustrations that many physicists and cosmologists have concerning
the status of inflation as sound science, and to provide another
perspective with which to continue the productive debate on the
subject.\footnote{
  With apologies to our colleagues, given the nature of this
  document and the familiarity of the community with the subject
  matter, we have included no references.
}

\subsection{An Allegory}

Is the inflationary paradigm good science?  By this we mean is it
falsifiable?  Are there any principles or predictions that are
inviolable?  These are the stringent questions that must be asked of
any scientific paradigm.  But, it is worth contemplating an analogy
before denouncing inflation.

Particle physics lays claim to a remarkable theoretical foundation,
its Standard Model.  This model, approximately thirty years old, has
been tested to an exquisite degree and, by the standards of cosmology,
holds up incredibly well.  But, just as one can ask whether the
inflationary paradigm is good science, one can as easily ask the same
of the Standard Model.  More accurately, we should ask whether the
gauge principle is good science.  Here, we view the gauge principle as
the governing concept that all fundamental interactions are mediated
by vector bosons that are universally coupled to fermionic matter,
representing a perfectly respected gauge symmetry.  This gauge
principle arising out of quantum electrodynamics is the foundation for
the Standard Model.

But is the gauge principle falsifiable?  What firm predictions does it
make?  Just as for inflation, there are many models that are
consistent with the paradigm, many gauge groups that may be
considered, many variations on the theme.  In the real world, one must
take the rather cumbersome $SU(3)\times SU(2)\times U(1)$ gauge group
to explain all the data.  Indeed, if the data were to be different,
one would modify the gauge group or add more particles to explain
every anomalous feature.

One can take this analogy even further.  Taken in its simplest form,
the gauge principle has definite predictions.  It must have massless
gauge bosons for every gauge symmetry present.  And while this
prediction works extraordinarily well for electromagnetism, it doesn't
work for the nuclear forces.  The weak gauge bosons are not massless.
One cannot even directly observe the gluons.  Not all gauge symmetries
are explicitly, or even approximately, respected.

Direct predictions of the simplest manifestation of the gauge
principle are categorically refuted by observation.  A whole new
system needs to be manufactured.  Outrageous modifications are made to
the gauge paradigm such as the addition of a fundamental scalar Higgs
boson with non-universal couplings and spontaneous symmetry breaking;
and a non-perturbative realization of the gauge principle must be
introduced for the color force in order to bring the gauge principle
into line with observations.  Should one then argue that the gauge
principle is garbage?  That it isn't science because one can modify it
{\it ad infinitum} in order to fit, however awkwardly, with the data?
And yet the gauge paradigm is considered wildly successful.  Why?  Is
this situation different from inflation?

\subsection{Lessons}

Particle physicists would be reluctant to characterize to the gauge
principle as lacking the heft of real science, or being totally devoid
of inviolable predictions, and therefore not falsifiable.  The answers
to the provocative questions raised above are that, indeed, the gauge
principle does have a set of inviolable principles: an exact (but
possibly hidden) gauge symmetry, gauge bosons mediating the associated
interactions, and universal couplings of those gauge bosons to matter.
Each of these predictions is indeed confirmed by observation.  All
variants of the Standard Model, however baroque, must respect these
principles.

In order to put inflation on the same footing as the gauge principle,
we need to enumerate a similar set of inviolable principles.  Put
another way, we need to identify what makes inflation so appealing
that it may suffer many alterations.  What are its inviolable
predictions?  What are its core principles?  The frustration with
inflation stems from the apparent scarcity of inviolable
principles, thanks to the ingenuity of creative inflationary theorists,
and the apparent scarcity of independent experiments with which
to test the self-consistency of inflation in the conceivable future.

\section{The Inflationary Paradigm}
\label{paradigms}

What are the principles underlying an inflationary theory?

We start by defining the {\em classical inflationary paradigm} as:
accelerated expansion of an initially marginal super-horizon (or,
super-Planck, if at $t=0$) volume, proceeding for many doubling times
(order 100), and ending everywhere (or at least over an exponentially
larger super-horizon volume) with thermalization and baryogenesis
(sufficient for successful nucleosynthesis).  Implicit in this
paradigm is some driving mechanism for the accelerated expansion and
the appropriate initial conditions that would lead to it.  In all
realizations of which we are aware, the driving mechanism is some
field, usually referred to as the ``inflaton.'' Suitable initial
conditions for the inflaton are assumed, usually on the basis that all
possible initial conditions are statistically realized. General
relativity (GR) is taken to be the dynamics of spacetime.

This classical paradigm, which arises out of classical field theory,
must be promoted to a quantum paradigm.  So long as we are interested
in spacetime curvature scales much less than the Planck scale, we
continue to treat gravity as classical; however, the inflaton field
must be treated quantum field theoretically.\footnote{
Also implicit
has been a particular description of the vacuum state of the theory
(the Bunch-Davies vacuum), extending possibly to trans-Planckian
energy scales (and hence sub-Planckian length scales), although some
researchers have begun to explore the robustness of this framework.}
Here there are two levels of complexity which we denote the {\em
semiclassical inflationary paradigm} and the {\em quantum inflationary
paradigm}.

In the semiclassical inflationary paradigm one is in the perturbative
regime of the quantum theory and quantum fluctuations can
self-consistently be regarded as occurring against a background of the
classical evolution of the inflaton field and the metric, at least
over a range of length scales extending up beyond our current Hubble
volume.  This paradigm is the one appropriate to new inflationary and
natural inflation models.  Moreover, it is this paradigm that is in
play whenever predictions of inflation are compared to observational
data.
 
In the quantum inflationary paradigm, for at least some portion of the
inflationary epoch, one is in the regime where backreaction of quantum
fluctuations on the spacetime need to be taken into account. To do
this properly, one would need to extend GR to include quantum
effects. This paradigm is the one appropriate to eternal, stochastic
or chaotic inflation.  The progenitors of inflation have argued that
the quantum paradigm is the most satisfying realization of the
inflationary paradigm, especially to alleviate the tuning of initial
conditions necessary to start inflation.  In this scenario the
Universe is bubbling with regions that are inflating. Inflation never
ends everywhere; nevertheless, there are pockets that stop inflating
and subsequently thermalize.  According to this quantum scenario, we
live in one of the thermalized regions.  Inflation is therefore
anthropic. Conditional probabilities for predictions are found with
the condition that the thermalized region be inhabitable. The Universe
is not homogeneous on the largest scales, but regions large enough to
accommodate our visible Universe can be smooth enough.

The apparent simplicity of these paradigms makes inflation so
attractive.  Unfortunately, it also means that there are only a few
generic features to characterize inflationary models of the Universe
observationally or experimentally.  Nevertheless, even these few
ingredients do seem to have certain consequences:

\subsection{Homogeneous, Isotropic Entropy-Filled Universe.}

That the accelerated expansion of the Universe ends everywhere is
implicit in the semiclassical paradigm. That it does so in the 
quantum paradigm is no less true, but much more subtle, 
incorporating generically the simultaneous truths that at any given 
place it eventually ends, but that it never ends everywhere, and the 
volume of space in which it has ended is vastly smaller than the volume 
in which it has not.  Indeed, in this picture it is often 
justified only anthropically why we do not inhabit a still-inflating region.  
Either way, we apparently must live in a region where the energy stored in 
the field driving inflation, the inflaton, was converted into other more 
prosaic forms of energy.  The vast amount of inflationary expansion is 
followed in all generic models by a rapid injection of entropy and its 
thermalization (through either reheating or preheating). This is taken 
to be governed by a Lagrangian density which is independent of space-time 
location.  It is difficult to put any measure on the predicted efficiency 
of this process, but the reheat temperature must be high enough to
allow nucleosynthesis.

In the semiclassical paradigm, the vast inflationary expansion
provides (almost) homogeneous initial conditions for entropy injection
and thermalization over some large length scale. This scale may
however be limited (as in $\lambda \phi^4$ theory) where, despite weak
coupling ($\lambda\ll1$) the semiclassical approximation
($\delta\rho/\rho\ll1$) fails on sufficiently large scales. Thus,
homogeneity sufficient to accommodate the semiclassical
assumption over a moderate range of scales is a consequence of weak
coupling.

In the quantum paradigm homogeneity seems to be an assumption 
that can be made self-consistently rather than a prediction. In this 
scenario, quantum fluctuations can be large, though inflation might 
be quenched wherever this happens.  Models exist in which the fluctuations
remain tamed.

\subsection{Super-Horizon Fluctuations}

The inevitable quantum fluctuations in the inflaton field will be
stretched beyond the cosmic horizon and imprint themselves in the
resulting energy density after reheating.  Only after inflation stops
and conventional big-bang evolution occurs will scales that left the
horizon during inflation reenter the cosmic horizon.  These
fluctuations thus appear super-horizon in scale.  Unfortunately,
there is no minimum predicted amplitude of scalar fluctuations;
their spectrum is model-dependent.

The same type of fluctuations would be produced for any light
(compared to inflationary Hubble scale), non-conformally-coupled
field, {\it e.g.}, gravity waves.  As with inflaton fluctuations,
these field fluctuations will be super-horizon. However, unlike the
inflaton, these fields are not expected to carry the bulk of the
Universe's energy density, and sifting for these particular signature
fields may be challenging.  The amplitude for super-horizon tensor
(gravity-wave) fluctuations is constrained from below by the
requirement that the post-inflationary reheat temperature be larger
than that necessary for nucleosynthesis.  In principle this constraint
offers a strictly falsifiable prediction of the inflationary paradigm,
though in practice the minimum amplitude is inaccessible for the
foreseeable future.

\subsection{Other Model-Independent Predictions}

Of course, there are other predictions, such as the existence of
inflaton particles that should appear at the inflationary mass scale.
These particles, however, may be extremely weakly coupled to
conventional matter and may be difficult to observe, even if one had
access to such energies.  Nevertheless, the inflaton field cannot be
completely decoupled from standard model physics.  A significant
amount of reheating to conventional particles requires some amount of
coupling.  This coupling may in principle be exploited, putting
inflation strictly within the regime of particle physics, and
providing another avenue for the falsification of inflation.
Unfortunately, unless the inflation energy scale is very low compared
to the Planck scale ({\it e.g.} near energies of $\approx 1~{\rm
TeV}$), this also remains an inaccessible possibility for the
foreseeable future.

\section{Model-Dependent Predictions}

Unfortunately, other predictions depend on the particular inflation
model employed.  As indicated earlier, the ingenuity of theorists
has shown that the idea that the Universe can be homogenized with an
early stage of accelerated expansion may be incorporated (with varying
degrees of ease) in an overwhelmingly diverse set of models.
However, we may take the predictions made by the simplest models as a
guide for what is more or less natural in an inflationary model.  

We can imagine a scenario where hypothetical observers know very
little about observational cosmology except that the Universe is very
old and filled with matter.  However, they have a great deal of
understanding about the rest of physics, and in particular, have been
led to believe that gravitation is intimately connected with the
dynamics of spacetime and that GR should govern the evolution of the
Universe.  In so doing, they would have realized, as have we, that the
age of the Universe, as determined from the ages of planetary and
meteoroidal material, is much greater than the only natural time scale
of GR -- the Planck time,  and that the curvature scale of the
Universe is much greater than the only natural length scale in GR --
the Planck length.  They might also have wondered where all the
entropy in the Universe came from, and why, in particular, the total
energy of everything they could see was much greater than the only
natural mass scale in GR -- the Planck mass.

Faced with these problems -- the age problem, the flatness problem and
the entropy problem -- they might well have developed the beautiful
paradigm of inflation: the idea that there was in the early history of
the Universe an epoch of accelerated expansion driven by the energy
density and negative pressure of the instantaneous vacuum state, which
serves to flatten the Universe, vastly increase the characteristic
dynamical time scale of cosmology, and fills the Universe with a
relatively homogeneous and abundant ``soup'' of particles.  In the
absence of any substantial data, physicists in this world would turn
to the most basic models of inflation to ascertain possible new
predictions about cosmology.

The simplest versions of inflation involve a single scalar field,
minimally coupled to gravity, with a potential polynomial in the
field, {\it e.g.}, $V = \lambda\phi^4$ where $\phi(x^\mu)$ is the
inflaton field, and $\lambda$ is small.  The inflaton begins trapped
in some state far away from the true vacuum, $\phi(t=0,{\bf x}) =
\phi_0 \gg M_P$, where $M_P$ is the Planck mass.  If $\phi_0 \ll
M_P\lambda^{-1/6}$, we are in the semiclassical paradigm.  The
inflaton field rolls slowly down the potential as the Universe engages
in accelerated expansion.  Eventually, the field exits the slow-roll
regime and coherently oscillates around the vacuum.  This oscillation
induces preheating and reheating to standard model particles, and the
Universe subsequently evolves via a standard hot-big-bang model.  If
$\phi_0 \gg M_P\lambda^{-1/6}$, we are in the quantum paradigm.  The
Universe begins in a stochastically inflating state but eventually
transitions into a regime where $\phi(t) \ll M_P\lambda^{-1/6}$ in some
region; semiclassical behavior subsequently dominates.  Evolution in
this region proceeds as in the semiclassical paradigm.

\subsection{Flat Universe}

In this simplest model, inflationary expansion flattens the Universe
beyond the ability of any likely experiment to discern a non-zero
value for $\vert\Omega-1\vert$.  Thus, the hypothetical cosmologists
would conclude that $\vert\Omega-1\vert$ should be so small as to not
be easily measurable.

One can see how this prediction can be easily avoided by looking
beyond the simplest models.  The original terrestrial 
(\lq\lq{old}\rq\rq) inflationary models, in which inflation ended 
via a first-order phase
transition generically predicted that if we live in a single bubble of
the true vacuum then the space-like hypersurfaces of constant
curvature should be hyperbolic ($\Omega < 1$).  (When cosmological
data suggested that indeed $\Omega\simeq0.3$, this fact was used to
argue that $\Omega\simeq 0.1-1$ was generic.)  However, first-order 
inflation (unless dressed up with double inflation,
topologically-non-trivial manifolds, or other complexifications) fails to
solve the suite of inflation-motivating cosmological problems.
Moreover, if even in the simplest models, inflation can accommodate
observably non-flat universes by allowing inflation to turn off at exactly
the correct number of e-foldings.  Of course, this just-so possibility is
often viewed as unpalatable and unnatural.

\subsection{$\delta\rho/\rho \lsim 1$}  

In the semiclassical regime, for an inflationary field $\phi$ with a
self-interaction potential $V(\phi)$, the amplitude of scalar
fluctuations (as opposed to vector or tensor modes) is
\begin{equation}
\label{drhobyrho_a}
\frac{\delta\rho}{\rho} \sim \frac{V^{3/2}}{V' M_{Pl}^2} \ .
\end{equation}
Specifically, to find the amplitude of fluctuations on a particular
scale, we evaluate the right-hand side of Eq.~(\ref{drhobyrho_a}) at
the value which $\phi$ held when that particular scale crossed out of
the apparent horizon.  It might seem that this easily could be much
less than unity.  However, during slow-roll, there is a relationship
between $\phi$ and the number of e-foldings until the end of
inflation, $N$,
\begin{equation}
N \sim \frac{\phi^2}{M_{Pl}^2}\ .
\end{equation}
For the model $V(\phi)=\lambda \phi^4$, Eq.~(\ref{drhobyrho_a}) may
be recast as
\begin{equation}
\label{drhobyrho_b}
\left.\frac{\delta\rho}{\rho}\right|_k \sim \lambda^{1/2} N_k^{3/2} \ ,
\end{equation}
where ${\delta\rho}/{\rho}|_k$ is the scalar fluctuation amplitude of
a given comoving wavenumber, $k$, where $N_k$ is the number of
e-foldings between when that scale left the inflationary horizon and
the end of inflation.  Those scales where ${\delta\rho}/{\rho}|_k >
{\cal O}(1)$ actually probe the stochastic regime of the quantum
inflationary paradigm, implying Eq.~(\ref{drhobyrho_b}) is no longer
valid.

We observe density fluctuations in the Universe over a given range of
comoving scales $k$ whose $N_k \sim 100$.
Equation~(\ref{drhobyrho_b}) then implies that even though $\lambda$
may be small enough for weak-coupling to be self-consistent, density
fluctuations need not be small.  For ${\delta\rho}/{\rho}|_k \ll 1$,
$\lambda$ must be further fine-tuned; the smaller the observed
fluctuations, the more fine-tuned $\lambda$ must be.  Alternatively,
one may venture into so-called natural inflation models, which exploit
almost-symmetries (such as pseudo-goldstone modes or flat directions
in dynamically broken supersymmetry) to explain unexpectedly small
density perturbations.

\subsection{Adiabatic fluctuations}

Because the energy in the field driving inflation is eventually 
converted into the thermal soup of radiation and matter filling the 
Universe, the inflaton would be converted into fluctuations in the 
cosmic energy density, and thence, through the dynamical response of 
the local geometry, into fluctuations in the metric, as well as the 
large-scale statistical distribution of matter in the Universe.  
Thus, the fluctuations would generically be adiabatic. In more
complicated models of inflation, however, the fluctuations can
have a non-adiabatic component.

\subsection{Gaussian fluctuations}

In the simplest inflationary models, the fluctuations arise from the 
excitation of independent inflaton modes. Therefore the statistics of 
each mode would be that of a Gaussian random field. In more complicated
inflationary models, it is seen that there can be small departures from 
Gaussianity.

\subsection{(Very Nearly) Equal Power on All Scales}

Equation (\ref{drhobyrho_a}) shows that $\delta\rho/\rho$ is a
function only of $V$ and $V'$.  Since to realize a large number of
e-folds of expansion $V(\phi)$ must be very flat, therefore the
amplitude of fluctuations generated on all scales should be nearly
equal.  The hypothetical cosmologists would therefore conclude that
the spectrum of fluctuations should be scale-free or very nearly so.
In particular, unless the scale corresponding to the onset of
inflation, or some other transitory event, just happens to have been
stretched to a physically observable scale -- less than the current
horizon size yet larger than the scale on which non-linear dynamics
confuses the traces of the primordial fluctuations -- there should be
no observable features in the primordial power spectrum that they
would deduce when they some day make measurements of structure beyond
their planetary system.

However, the detailed structure of the power spectrum
depends on the exact form of the inflaton potential, the potential can
be tuned in such a way as to provide whatever power spectrum is
necessary, within some broad constraints that slow-roll inflation
require.  It is no wonder why many cosmologists invariably
point to this feature of inflation and regard it as dangerously
epicyclic.  That one can tune the spectrum with an arbitrarily pliable
inflaton potential to fit most any given spectrum is disturbingly
unsatisfying.

\section{Observations and Evaluation of the Paradigm}

In our hypothetical scenario, eventually observational cosmology as we
know it would be revealed.  We here summarize current observations,
and evaluate the inflationary paradigm in light of each piece of
evidence.

\begin{enumerate}

\item {\bf A homogeneous, full Universe}.
Measurements of the CMB probe primarily our past light cone, 
and mostly the surface of last scattering. Although Occam's razor 
suggests that it is highly unlikely that we just happen to live at 
the center (within parts per billion by volume) of a spherically 
symmetric inhomogeneous universe, direct observational probes of the 
interior of the light cone are harder to come by. However, observations 
of distant galaxies 
establish that element abundances are uniform across the Universe, 
suggesting that there were no large fluctuations in the energy 
density or baryon number at the time of primordial nucleosynthesis.  
Having had ample time to investigate the details of the onset and
dynamics of inflation, we may well be reluctant to claim that the 
homogeneity and isotropy of the Universe
are really great successes of inflation since the onset of inflation
in any particular patch of space requires that that patch be
relatively homogeneous on super-horizon scales to begin with (although
once it is, inflation can vastly improve the homogeneity). Moreover,
other theories (such as variable speed of light and various braneworld
scenarios) may also explain the homogeneity and isotropy, so these
features are not terribly good discriminators between theories.
Consistent with the classical paradigm, the
visible Universe has a very large entropy, $S\simeq 10^{87}$.

\item {\bf Super-horizon fluctuations}.
The observation of acoustic peaks in the angular power spectrum of 
the CMB and in particular, as discussed by the WMAP 
team, the anti-correlation between the temperature 
anisotropy and the E-mode polarization at $1-2^\circ$ angular scales 
establishes that super-horizon scalar fluctuations exist.  This 
observation is a true cause of celebration for the inflationary 
paradigm. While other theories may also predict such fluctuations, 
they really are a generic feature of all inflationary models.  Tensor
fluctuations have not yet been observed.  This sets a mildly
interesting limit on the inflationary energy scale, but of course
far above the minimum energy scale required by nucleosynthesis.

\end{enumerate}

The rudiments of the inflationary paradigm seem to hold up to
scrutiny.  However, as we commented, these are extremely
limited, and lack a great deal of discriminatory power.  What of
predictions of the simplest models?  How surprised would our
hypothetical cosmologists be?

\begin{enumerate}

\item  {\bf A flat Universe}.
The discovery and clear definition of the first peak
in the angular power spectrum of the cosmic microwave background (CMB)
established definitively that the Universe is
flat or nearly so, $\Omega\simeq 1$, in particular that $\Omega\neq
0.3$ as had previously been widely considered.  Analysis of the WMAP
observations show that $\Omega=1.02\pm0.02$. In the simplest models 
of inflation, which are the only ones we are considering here, 
$\Omega$ is predicted to be unity to very high precision. This 
fits in well with observation.

\item {\bf An extremely homogeneous Universe}.
The original discovery of the $2.7~{\rm K}$ CMB radiation by Penzias
and Wilson in 1965, was soon followed by efforts
to measure any anisotropy in that background.  However, it was not
until 1991 that the first successful measurement of the anisotropy was
made by the Cosmic Background Explorer satellite. The long
delay was due to the very small amplitude of the anisotropy, only
parts per $10^5$.  Since then many experiments have measured this
anisotropy and its properties.  The fine-tuning needed to achieve the
observed $\delta\rho/\rho$ would concern our hypothetical
cosmologists.  Because we developed inflation with the foreknowledge
that $\delta\rho/\rho \ll 1$, we have been more prepared to accept
{\it a priori} this fine-tuning problem.  As
limits on $\delta\rho/\rho$ improved through the 1980's, the
fine-tuning grew ever more severe, but it did so adiabatically,
forestalling any increasing sense of concern.

\item {\bf Adiabatic fluctuations}.
All known observations are consistent with all fluctuations being 
entirely adiabatic in nature.  As reported by WMAP 
the fit to their data is not improved by adding any amount of 
isocurvature fluctuations. This is good support for acausal generation 
of perturbations, and fits in very well with the simplest models
of inflation. So the consistency of adiabaticity is important, 
but the limits on non-adiabaticity remain weak. Also a number of 
inflationary models have been constructed that generate non-adiabatic 
fluctuations. 

\item {\bf Gaussian fluctuations}.  No deviations from Gaussianity
have been observed in the fluctuation spectrum. The absence of any
detected non-Gaussianity of the fluctuations would likely be viewed as
a relief, but hardly a coup, since very nearly Gaussian distributions
are rather generic due to the central limit theorem; moreover, unless
one knows what non-Gaussianity to look for, finding it is really like
finding a needle in a very large haystack.  It is again to be
noted that there exist several inflationary models that predict
non-Gaussian fluctuations.

\item {\bf Lack of equal power on all scales}.  On scales
characterized by $\ell$s from ten to several hundred, the angular
power spectrum, as determined by many CMB experiments, and
particularly by WMAP is nearly scale free.  However, COBE-DMR found
and WMAP has confirmed that on angular scales greater than about
$60^\circ$, the two point angular correlation function of the CMB
temperature fluctuations nearly vanishes.  The WMAP team has argued
that the best fit standard $\Lambda{CDM}$ model is ruled out at the
99.85\% confidence level based on comparisons of the observed
$C(\theta)$ with a Monte Carlo of $10^5$ realizations of the model.
Mild adjustments of the model only improve that to a 99.7\% exclusion.
The absence of these correlations on large angular scales is a serious
problem for inflation. This is not because there exist no inflationary
models which accommodate it.  Features in the inflaton potential, two
stage inflation, just-so inflation in a compact manifold, braneworld
models, {\it etc.} all may hold promise of accommodating this data.
However, unless such modifications offer additional testable
predictions, they are, indeed, dangerously epicyclic.

\end{enumerate}

\section{Concluding Remarks}

Post-WMAP statements have been made claiming that the predictions of
inflation have been confirmed, and that inflation is a successful
paradigm.  However, careful consideration of the meaning of the term
``inflationary paradigm'' suggests that such statements are, at best,
imprecise.  Generic predictions of the {\em inflationary paradigms}
depend on certain assumptions that are rarely made explicit.  Granting
these assumptions, the essential predictions of homogeneity and
isotropy, and the existence of super-horizon fluctuations are indeed
confirmed; however, only the latter is a post-inflation discovery.

Further implementation of the inflationary paradigm requires adopting
a particular model.  The simplest one-field inflation models have
met with limited success when confronted by new data.  The Universe
appears spatially flat, and fluctuations are adiabatic and Gaussian.
However, the fluctuation amplitude is unnaturally small and are
decidedly not scale-free on the largest angular scales.  While the
former requires only a fine-tuning of Lagrangian parameters, the
absence of large-scale power seems to demand models that are
carefully designed.  But, unless these new models yield testable
predictions, this tack merely perpetuates the habits that inflation's
critics abhor.  Do we continue to accept
inflation merely because there is no better alternative?

Science is not a democratic pursuit.  It only takes one contradictory
fact to consign a theory to the dustbin of history, or at least to
take it off its pedestal and send it back to the workshop.  On the
other hand, when one poses a given paradigm, it always make sense to
begin with the simplest incarnation of that paradigm.  The degree to
which a model must be engineered to reproduce the needed data should
then be factored into a reassessment of the worth of the original
idea.  If a theory is repeatedly faced with contradictory facts which
force a reengineering, at what point does it stop being good science?
If this is to be the dawn of a new era of precision cosmology, it must
involve not only precise determinations of an ever increasing number
of new parameters, but also precision tests of the self-consistency of
our theories which permit their dispassionate evaluation.

\acknowledgements

We are grateful to Serge Winitzki for clarifying remarks on the 
quantum inflationary scenario.  This work is sponsored by DOE
Grant DEFG0295ER40898 and CWRU Office of the Provost.

\end{document}